\documentclass{PoS}

\newcommand{\eq}[1]{eq.~(\ref{#1})}
\newcommand{\Eq}[1]{Eq.~(\ref{#1})}
\newcommand{\fig}[1]{Fig.~\ref{#1}}

\newcommand{\lag}[1]{{\mathcal{L}}_{\rm {#1}}}
\newcommand{\tr}{\mathrm{tr}\,}

\newcommand{\ev}[1]{\left\langle #1 \right\rangle}
\newcommand{\gbar}{\bar{g}}
\newcommand{\mbar}{\kern1pt\overline{\kern-1pt m\kern-1pt}\kern1pt}
\newcommand{\msbar}{{\rm \overline{MS\kern-0.05em}\kern0.05em}}
\newcommand{\SU}[1]{SU(#1)}
\newcommand{\su}[1]{\mathfrak{su}(#1)}
\newcommand{\gauge} [3][U]{\ensuremath{#1_{#2}(#3)}}
\newcommand{\plaq} [4][P]{\ensuremath{#1_{#2,#3}(#4)}}
\newcommand{\Sw}{S_{\rm w}}
\newcommand{\rD}[1]{{\mathcal D}[#1]}

\def\Dw{D_{\mathrm{w}}}
\def\nf{N_{\mathrm{f}}}

\def\rb{r_{\mathrm{b}}}
\def\CF{C_{\rm F}}
\def\gqq{\bar{g}_\mathrm{qq}}
\def\aqq{\alpha_{\rm qq}}
\def\aqqr{\alpha_{\rm qq}(1/r)}
\def\betaqq{\beta_\mathrm{qq}}
\def\bqq{b^{\rm(qq)}}
\def\gc{\bar{g}_\mathrm{c}}
\newcommand{\www}{\mbox{
\begin{minipage}{24pt}\includegraphics[width=24pt]{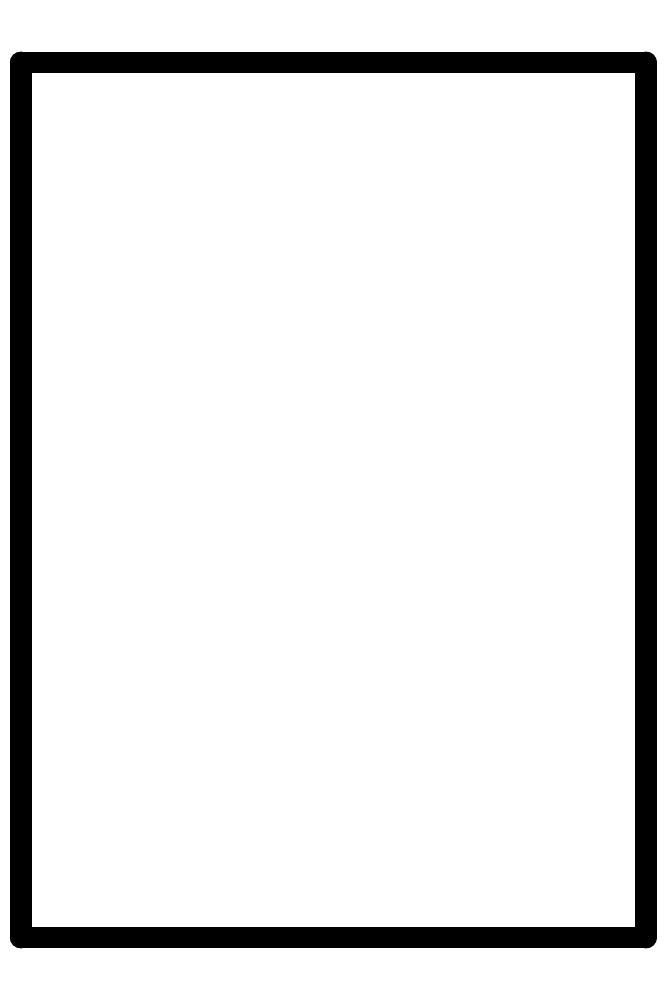}\end{minipage}}}
\newcommand{\wbbc}{\mbox{
\begin{minipage}{24pt}\includegraphics[width=24pt]{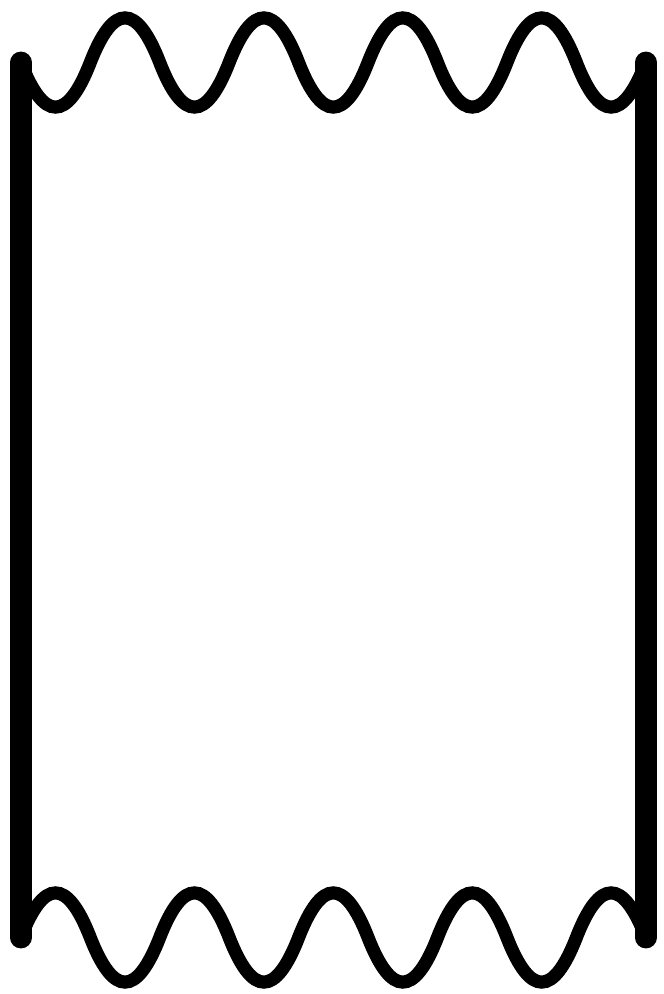}\end{minipage}}}
\newcommand{\wbbd}{\mbox{
\begin{minipage}{24pt}\includegraphics[width=24pt]{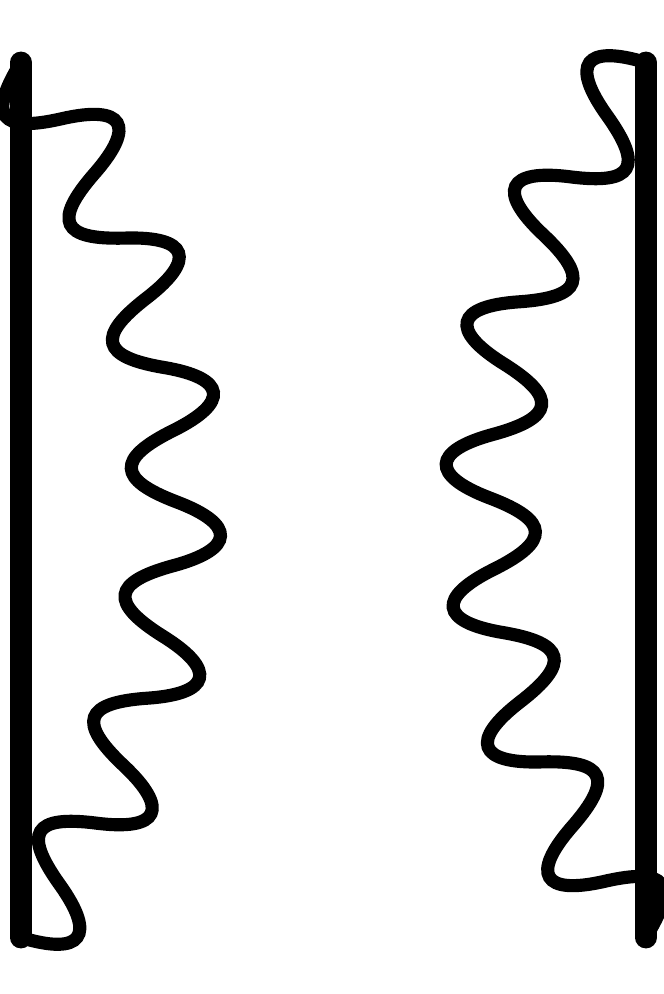}\end{minipage}}}
\newcommand{\wwb}{\mbox{
\begin{minipage}{24pt}\includegraphics[width=24pt]{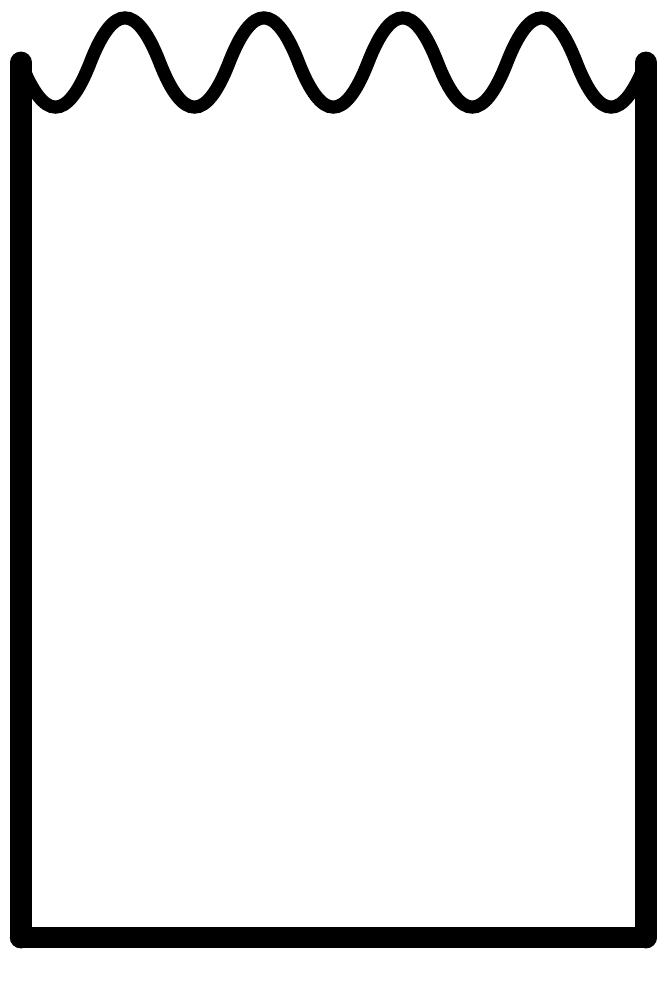}\end{minipage}}}
\newcommand{\wbw}{\mbox{
\begin{minipage}{24pt}\includegraphics[width=24pt]{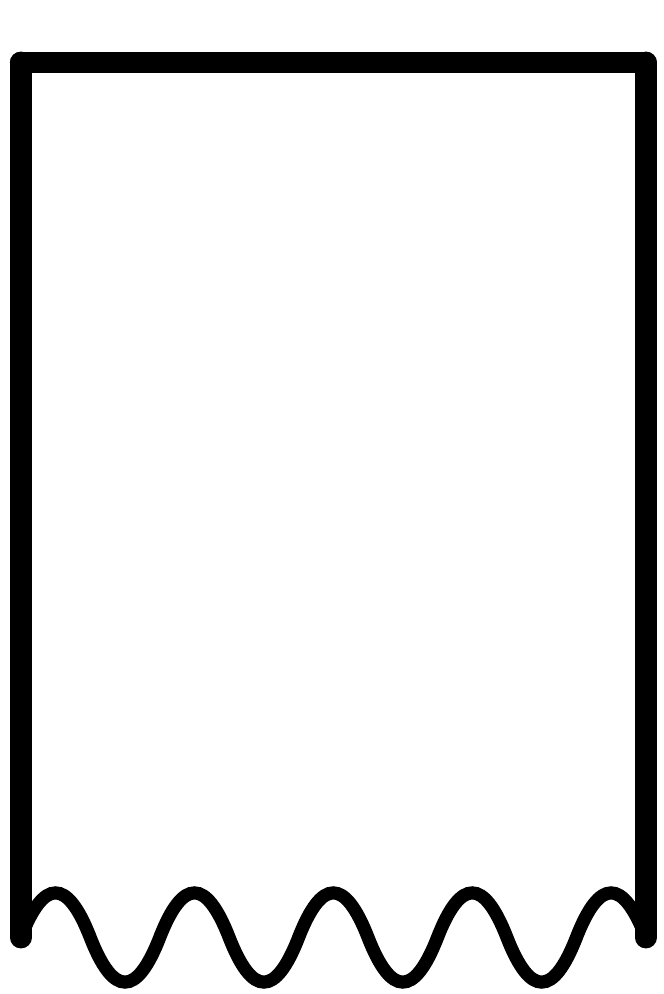}\end{minipage}}}

\def\as{\alpha_{\rm s}}
\def\ac{\alpha_{\rm c}}

\title{Lattice Quantum Chromodynamics}

\ShortTitle{Lattice QCD}

\author{\speaker{Francesco Knechtli}
        \\
        Department of Physics, Bergische Universit\"at Wuppertal\\
        Gau\ss{}stra\ss{}e 20, 42119 Wuppertal, Germany\\
        E-mail: \email{knechtli@physik.uni-wuppertal.de}}

\abstract{This lecture provides an introduction to quantum chromodynamics (QCD) on the lattice. The continuum limit and Monte Carlo simulations are briefly discussed. Different facets of QCD are nicely exhibited by the potential of a static quark and anti-quark pair and results from lattice calculations of this quantity will be presented.}

\FullConference{Corfu Summer Institute 2016 "School and Workshops on Elementary Particle Physics and Gravity"\\
		31 August - 23 September, 2016\\
		Corfu, Greece}

\begin{document}

\section{Lattice QCD}

\begin{figure}[t]\centering
\includegraphics[width=.7\textwidth]{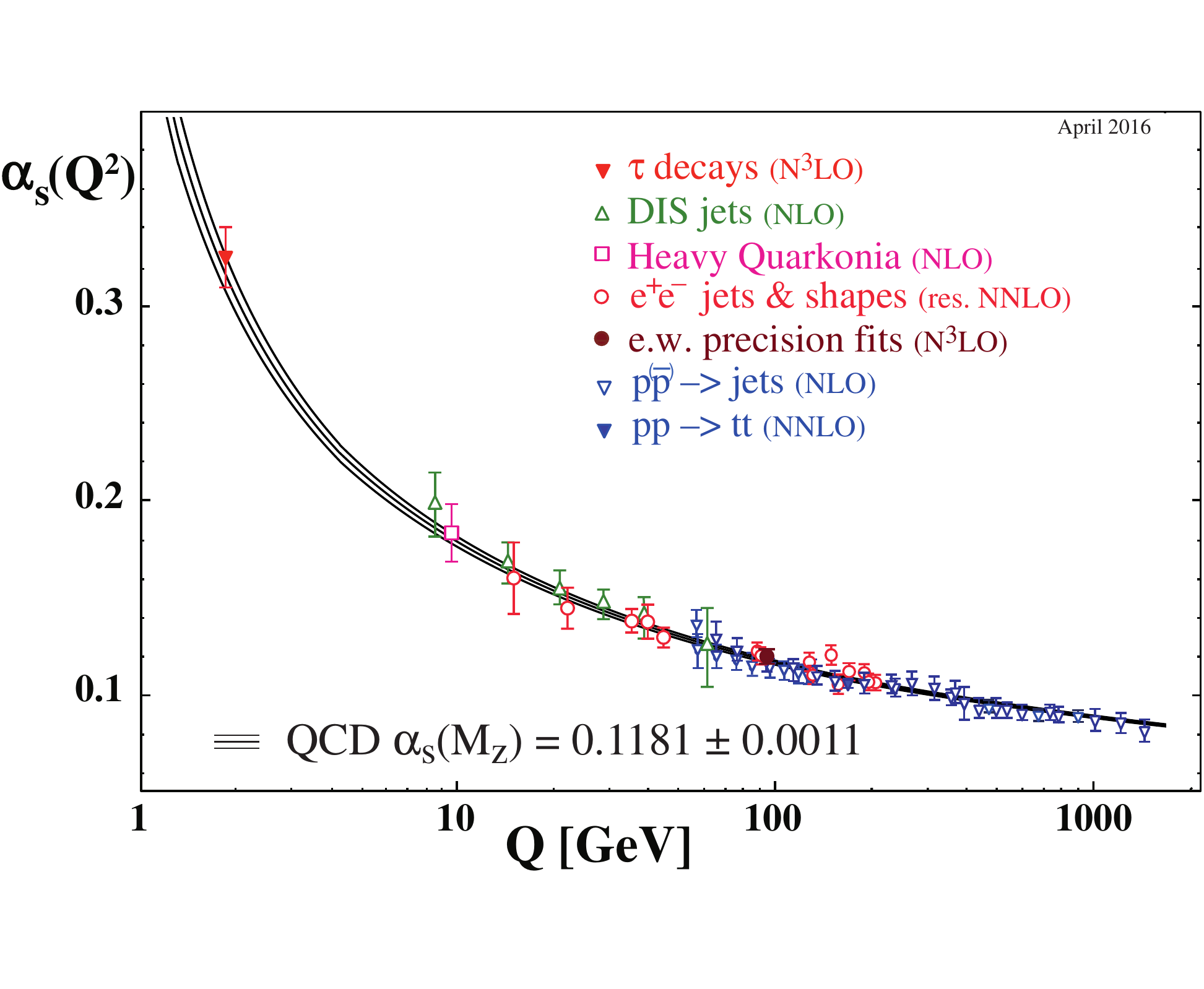}
\caption{The strong coupling $\alpha_{\rm s}(\mu)=\gbar^2(\mu)/(4\pi)$ in the $\msbar$ scheme taken from Ref.~\cite{Olive:2016xmw}.}
  \label{f:asq}
\end{figure}

This introductory section on lattice Quantumchromodynamics (QCD) will be brief.
More details can be found in \cite{Knechtli:2017sna}.
QCD is the theory of strong interactions with the Euclidean Lagrangian
\begin{equation}\label{e:lagQCD}
 \lag{QCD}(g_0,m_q) = -\frac{1}{2g_0^2}\tr\{F_{\mu\nu}F_{\mu\nu}\} +
 \sum_{q=u,d,s,c,b,t} \bar{q}\,(\gamma_\mu(\partial_\mu + A_\mu) + m_q)\,q
\end{equation}
The free parameters of the theory are the gauge coupling $g_0$ and the 
quark masses $m_q$.
For small $g_0$ calculations in QCD can be performed by an asymptotic expansion
in $g_0$ called perturbation theory. 
The interaction vertex of a quark, an anti-quark and a gluon is proportional 
at tree level to the gauge coupling $g_0$. When higher order effects in 
perturbation theory are included the strength of the interaction is given
by an effective coupling $\gbar(\mu)$ which depends on the 
magnitude $\mu$ of the energy-momentum transferred by the gluon to the quarks,
see \cite{Luscher:2002pz}.
Asymptotic freedom is the property that for large $\mu$ the coupling 
$\gbar(\mu)$ goes to zero as \cite{Gross:1973id,Politzer:1973fx}
\begin{equation}\label{e:afree}
  {\bar g}^2(\mu) \buildrel \mu\rightarrow\infty\over\sim
  \frac{1}{2b_0\log(\mu/\Lambda)} \,,
\end{equation}
where $b_0$ is a known coefficient and $\Lambda$ is a mass scale called the
$\Lambda$ parameter of QCD. Asymptotic freedom is verified 
experimentally as shown in \fig{f:asq}.

At zero temperature and zero chemical potential QCD has two energy regimes:
\begin{enumerate}
\item at high energy $\mu\gtrsim2\,\mathrm{GeV}$, $\alpha_{\rm s}(\mu)\equiv \gbar^2/(4\pi)$ is small and quarks and gluons are (almost) free;
\item at lower energies, quarks and gluons are confined into hadrons.
\end{enumerate}
While perturbation theory works in regime 1. we need a different tool to extract
the properties of hadrons from $\lag{QCD}$ (regime 2.). The latter is provided
by the lattice formulation which is amenable to computer simulations.

QCD can be formulated as a path integral. Wilson \cite{Wilson:1974sk} and
independently Smit \cite{Wilson:2004de}
formulated this path integral on a four-dimensional Euclidean hypercubic 
lattice with lattice spacing $a$. The points of the lattice are denoted by
$x=(n_0,n_1,n_2,n_3)a$, $n_\mu=0,1,\ldots$ and the four directions by 
$\mu=0,1,2,3$.
For simulations a finite lattice of size $T\times L^3$ is considered.
A $\SU{N}$ gauge field consists of matrices $\gauge{\mu}{x}\in\SU{N}$
associated with oriented links between two nearest-neigbor points on
the lattice. A local gauge transformation is a set $\Omega(x)\in\SU{N}$.
The gauge links transform as
\begin{equation}\label{e:link_gtrsf}
\gauge[U^{\mathrm (g)}]{\mu}{x} = \Omega(x)\,\gauge{\mu}{x}\,\Omega^{-1}(x+a\hat{\mu}) \,.
\end{equation}
Integration over link variables is performed using the
group invariant (Haar) measure ${\rm d}U$:
\begin{equation}\label{e:Haar}
I[f] = \int_{\SU{N}}{\rm d}U\,f(U) = \int_{\SU{N}}{\rm d}U\,f(VUW)\,,\quad I[1]=1 \,,
\end{equation}
where the matrices $V$ and $W$ are in $\SU{N}$.
Plaquettes are the elementary (Wilson) loops on the lattice which are defined by
\begin{equation}\label{e:plaq}
\plaq{\mu}{\nu}{x} = 
 \gauge{\mu}{x}\,\gauge{\nu}{x+a\hat{\mu}}\,\gauge[U^{-1}]{\mu}{x+a\hat{\nu}}\,\gauge[U^{-1}]{\nu}{x}
\end{equation}
with the property $\plaq[P^{-1}]{\mu}{\nu}{x} = \plaq[P^{\dagger}]{\mu}{\nu}{x} = \plaq{\nu}{\mu}{x}$. Under gauge transformation the plaquette transforms as
$\plaq[P^{\mathrm (g)}]{\mu}{\nu}{x}=\Omega(x)~\plaq{\mu}{\nu}{x}~\Omega^{-1}(x)$.
Wilson's gauge action is defined by
\begin{equation}\label{e:Swgauge}
\Sw[U] = \frac{\beta}{N} \sum_{x}\sum_{\mu<\nu} 
 {\rm Re}\,\tr[1 - \plaq{\mu}{\nu}{x}] \,.
\end{equation}
It is positive, gauge invariant and has a parameter $\beta\ge0$.
The latter is related at the classical level to the continuum gauge
coupling $g_0$ through the relation
\begin{equation}\label{e:betapar}
 \beta = \frac{2N}{g_0^2} \,.
\end{equation}
The Euclidean path integral on the lattice is formulated as a statistical
mechanical system with partition function
\begin{equation}\label{e:Zgauge}
Z = \int\! \rD{U} \;\; {\rm e}^{-\Sw[U]}\,,\quad
\rD{U}=\prod_{x,\mu} {\rm d}\gauge{\mu}{x}
\end{equation}
with a compact Haar measure. This is a non-perturbative definition of the
Euclidean path integral.
An observable is a function of the gauge field ${\cal O}[U]$ and its
expectation value is
\begin{equation}\label{e:evalue}
\langle {\cal O} \rangle = \frac{1}{Z}
\int\! \rD{U} \;\; {\cal O}[U] \; {\rm e}^{-\Sw[U]} \,.
\end{equation}
Gauge fixing is not required for gauge invariant observables.

Quark fields $\psi_i^\alpha(x)$ are Grassmann variables with indices
$\alpha=1,\ldots,4$ (Dirac) and $i=1,\ldots,N$ (gauge; 
fundamental representation).
Antiquark fields $\bar{\psi}_i^\alpha$ are independent Grassmann variables
in the complex conjugate representation.
They obey the anticommutation relations
\begin{equation}\label{e:quarks}
\{ \psi^\alpha_i(x) , \psi^\beta_j(y) \} = 0 \,,\quad
\{ \bar{\psi}^\alpha_i(x) , \bar{\psi}^\beta_j(y) \} = 0 \,,\quad
\{ \psi^\alpha_i(x) , \bar{\psi}^\beta_j(y) \} = 0 \,.
\end{equation}
Under local gauge transformation quark fields transform as
\begin{equation}\label{e:quarks_gtrsf}
(\psi^{\mathrm{(g)}})_i^\alpha(x)=[\Omega(x)]_{ij}\psi^\alpha_j(x)
\quad\mbox{and}\quad
(\bar{\psi}^{\mathrm{(g)}})^\alpha_i(x)=\bar{\psi}_j^\alpha[\Omega(x)^{-1}]_{ji} \,.
\end{equation}
Wilson's action for the quark fields is defined by
\begin{equation}\label{e:Swquarks}
S_{\rm F} = a^4\sum_x \bar{\psi}_i^\alpha(x)[\Dw+m_0]_{ij}^{\alpha\beta}\psi_j^\beta(x)\,,
\end{equation}
where $m_0$ is the bare quark mass. The Wilson--Dirac operator is
\begin{equation}\label{e:Dw}
[\Dw]^{\alpha\beta}_{ij} = \frac{1}{2}\sum_\mu\left\{
[\gamma_\mu]^{\alpha\beta}(\nabla^*_\mu+\nabla_\mu)_{ij}
-a[\delta]^{\alpha\beta}(\nabla^*_\mu\nabla_\mu)_{ij}\right\}
\end{equation}
and contains the covariant difference operators
\begin{eqnarray}
(\nabla_\mu\psi)^\alpha_i(x) & = &
\{[\gauge{\mu}{x}]_{ij}\psi^\alpha_j(x+a\hat{\mu})-\psi_i^\alpha(x)\}/a \,, 
\label{e:covder_forward}\\
(\nabla^*_\mu\psi)^\alpha_i(x) & = & 
\{\psi^\alpha_i(x)
-[\gauge[U^{-1}]{\mu}{x-a\hat{\mu}}]_{ij}\psi^\alpha_j(x-a\hat{\mu})\}/a \,.
\label{e:covder_backward}
\end{eqnarray}
$\Dw$ is a large sparse matrix of dimension $12 T/a (L/a)^3 \times 12 T/a (L/a)^3$. 
The Nielsen-Ninomiya theorem states that
it is impossible to construct a local (free) lattice operator $D$ with 
$\{D,\gamma_5\} = 0$ without doubling the quark species.
The property $\{D,\gamma_5\} = 0$ guarantees the invariance under
(continuum) chiral transformations of the quark fields.
The term $-\frac{a}{2}\nabla^*_\mu\nabla_\mu$ in \eq{e:Dw} removes the
doublers (fermion copies) in the continuum limit but since 
$\{\Dw,\gamma_5\}=-a\nabla^*_\mu\nabla_\mu\neq0$ it violates chiral symmetry
at O($a$). By including the Sheikholeslami-Wohlert term in the Wilson--Dirac 
operator these violations can by reduced to O($a^2$) \cite{Sheikholeslami:1985ij,Heatlie:1990kg}.
A consequence of the breaking of chiral symmetry is that the bare quark
mass parameter $m_0$ in \eq{e:Swquarks} has to be tuned to a non-zero
negative critical value in order to realize zero physical quark mass.

A way to avoid the Nielsen--Ninomiya theorem goes back to an old suggestion
by Ginsparg and Wilson \cite{Ginsparg:1981bj}
by demanding only $\{D,\gamma_5\} = aD\gamma_5D$.
This idea was revived in 1997 \cite{Hasenfratz:1998ri,Luscher:1998pqa,Neuberger:1997fp}
and L{\"u}scher showed that there exists
a lattice form of chiral symmetry which is exactly preserved in this case. 

At this point we can formulate QCD on the lattice in terms of a gauge field
$\gauge{\mu}{x}\in\SU{3}$ and $\nf$ quark fields 
$\psi_f$, $f=1,\ldots,\nf$ with bare masses $m_{0f}$. The partition function is
\begin{equation}
Z = \int\! \rD{U}\;\; \prod_{f=1}^{\nf} \rD{\bar{\psi}_f}\,\rD{\psi_f}\;\;
{\rm e}^{-\Sw[U]+\sum_{f=1}^{\nf}\bar{\psi}_f(\Dw+m_{0f})\psi_f} \,,
\end{equation}
with $\rD{\psi_f}=\prod_{x,i,\alpha} {\rm d}(\psi_f)^\alpha_i(x)$. Using the
Matthews--Salam formula: $\int\!  \rD{\eta} \rD{\bar\eta} \;\;
 {\rm e}^{\bar{\eta}_iM_{ij}\eta_j}
 = \det(M)$
we arrive at the expression
\begin{equation}\label{e:ZQCD}
Z = \int\! \rD{U}\;\; {\rm e}^{-\Sw[U]}\,\prod_{f=1}^{\nf}\det(\Dw+m_{0f}) \,.
\end{equation}
A lattice operator for a nucleon at rest is
$N\sim\sum_{\underline{x}}\epsilon_{ijk}u(\underline{x})_i\left(u(\underline{x})_j^TC\gamma_5d(\underline{x})_k\right)$
in terms of up and down quark fields $u$ and $d$, see \cite{Gattringer:2010zz}. 
Assuming the existence of an (effective) transfer matrix\footnote{
Its existence for the lattice gauge theory with Wilson fermions and
Wilson plaquette action has been rigorously proven in \cite{Luscher:1976ms}.}
the nucleon two-point function
\begin{eqnarray}\label{e:N2pt}
\ev{N(t=na,\underline{p}=0)\overline{N}(0)} & \propto &
\int\! \rD{U}\,\rD{\bar{q}}\,\rD{q}\;\; {\rm e}^{-S[U,\bar{q},q]}\,
N(t,\underline{0})\overline{N}(0) \\
& \stackrel{\mbox{$n$ large}}{\sim} & {\rm e}^{-n/\xi_N(g_0,\ldots)}
\end{eqnarray}
decays exponentially with exponent $t\,m_N=n\,(am_N)$ where
$am_N(g_0,\ldots)=1/\xi_N(g_0,\ldots)$ is the nucleon mass in lattice units.
$\xi_N$ is the nucleon correlation length, which depends on the 
bare gauge coupling $g_0$ and other parameters like the quark masses.

\section{Continuum limit}

We consider for simplicity a QCD-like theory with massless quarks.
The input parameter is the lattice gauge coupling $\beta=6/g_0^2$.
The value of the lattice spacing $a$ is not an input. In order to
determine the lattice spacing (what is called scale setting, see
\cite{Sommer:2014mea} for a recent discussion) we 
can compute for example a mass, like the nucleon mass $am_N(g_0)$
and declare
\begin{equation}  
a(g_0) = \frac{am_N(g_0)}{m_N} \,,
\end{equation}
where $m_N$ is the physical value of the nucleon mass in units of $\rm{MeV}$.
Note that the value of $a$ depends on the chosen hadron mass.
Evenually we want to take the continuum limit $a\to0$. If we
consider the physical nucleon mass $m_N$ fixed this implies that 
the nucleon correlation length $\xi_N\to\infty$ diverges in the
continuum limit. The continuum limit can therefore be taken if
a critical value of the gauge coupling $g_{\rm crit}$ exists such that
$\xi_N\to\infty$ when $g_0\to g_{\rm crit}$.
The conventional wisdom is that
\begin{equation}\label{e:gcrit}
g_{\rm crit}=0 \,,
\end{equation}
in other words the continuum limit of lattice QCD is asymptotically free.
There is strong evidence in support of this statement (coming from
perturbation theory and simulation results) but it is not yet rigorously proven.
\Eq{e:gcrit} also holds when the quark masses are non-zero. In that case
the values of the bare quark masses are fixed by matching to experimentally
determined quantities.

Consider the mass spectrum of QCD $am_k$, $k=1,2,\ldots$ near the 
continuum limit.
Presently attained values of $\xi_k=1/(am_k)$ are not so large and one
needs extrapolations
\begin{equation}
\left[\frac{m_k}{m_1}\right](g_0) =
\left[\frac{m_k}{m_1}\right](g_{\rm crit}) + c_k {\rm O}((am_1)^p)\,, k>1 \,,
\end{equation}
where $p=1,2,\ldots$ and the powers of $a$ can be modulated by logarithmic
factors.
Symanzik's conjecture is that as $a\to0$ the lattice theory
can be described by an effective continuum theory with the lattice spacing $a$
as an expansion parameter \cite{Symanzik:1979ph}. Only local 
operators with the same symmetries as the 
lattice theory appear.
The expectations based on Symanzik's analysis are
\begin{itemize}
\item[a)] generic O($a^2$) ($p=2$) artifacts in pure $\SU{N}$ theory but
O($a$) ($p=1$) effects with pure Wilson fermions;
\item[b)] it is possible to construct improved lattice actions, for which
the leading cut-off effects are cancelled, in particular
O($a^2$)-improved lattice action for $\SU{N}$ theory and 
O($a$)-improved Wilson fermions.
\end{itemize}
We refer to a review by P.~Weisz \cite{Weisz:2010nr} for a review on 
Symanzik's effective theory of lattice artifacts.

\section{Monte Carlo simulations}

We restrict the discussion of Monte Carlo simulation to the
case of a pure gauge theory. The task is to compute expectation values of
functions of observables $O[U]$ defined by the integral in \eq{e:evalue}.
The dimension of this integral is
$(N^2-1)\times 4\times (T/a)\times (L/a)^3$, which can be O($10^9$).
It is not possible to solve it anlytically but it can be estimated by
a Monte Carlo simulation. The latter consists of generating a sequence
$\{U_1,U_2,\ldots, U_n \}$ (called an ensemble) drawn at random from the 
probability distribution $\Pi(U)={\rm e}^{-\Sw[U]}/Z$. 
The integral \eq{e:evalue} is then approximated by the ensemble average 
\begin{equation}\label{e:evalueestimate}
\ev{O} = \frac{1}{n}\sum_{i=1}^{n} O(U_i) + {\rm O}(n^{-1/2}) \,.
\end{equation}
Except for simple cases, it is not possible to generate independent
configurations in a sample. Instead one uses a Markov chain where $U_k$ is 
obtained from $U_{k-1}$ by a stochastic process. The chain depends on
$U_1$ and a transition probability function $T(U,U')$ with the
properties \cite{Luscher:2010ae}
\begin{enumerate}
\item $T(U,U')\ge0\;\forall U\,,U'$,~~~ $\int\; \rD{U'}\, T(U,U')=1\;\forall U$.
\item Stability: $\int\; \rD{U}\, \Pi(U)T(U,U') = \Pi(U')\;\forall U'$.
\item Aperiodicity: $T(U,U)>0\;\forall U$.
\item Ergodicity: for any given subregion of configuration space ${\cal S}$ 
one can find $U\in{\cal S}$ and $U'\notin{\cal S}$ such that $T(U,U')>0$.
\end{enumerate}
Usually it is easier to construct a transition function which fulfills the
detailed balance condition: $\Pi(U)T(U,U') = \Pi(U')T(U',U)$ for all pairs
$U'$, $U$. Detailed balance implies stability.
As an example, a simple algorithm to update gauge links in a Markov chain
is defined as follows:
\begin{enumerate}
\item Choose a link $(x,\mu)$ at random.
\item Choose $X\in\su{3}$ randomly in a ball $||X||\le\epsilon$ with 
uniform distribution, where $\epsilon>0$.
\item Generate a random number $r$ uniformly distributed in the interval 
$[0,1]$ and accept the new link $\gauge[U']{\mu}{x}={\rm e}^{X}\gauge{\mu}{x}$
if ${\rm e}^{-\Sw[U']}\ge r{\rm e}^{-\Sw[U]}$, 
otherwise keep the old link $\gauge{\mu}{x}$.
\end{enumerate}
For large-scale simulations we need excellent random number
generators, e.g RANLUX \cite{Luscher:1993dy}.
For the description of more efficient gauge link updates as well as
algorithms of full QCD including dynamical quarks we refer to
\cite{Knechtli:2017sna} and references therein.

In a Markov chain consecutive configurations are not statistically independent.
There are autocorrelations which effectively reduce the number of independent 
measurement of an observable $O$ by 
$n\longrightarrow n/(2\tau_{\mathrm{int}}(O))$, where $\tau_{\mathrm{int}}(O)$ is
called the integrated autocorrelation time. The latter depends on the algorithm
as well as on the observable, see \cite{Wolff:2003sm}.
Critical slowing down is the property that in the continuum limit
autocorrelation times diverge with some power of the inverse lattice spacing 
$\tau_{\rm int}(O)\propto (1/a)^{z(O)}$. The power $z(O)$ is called
dynamical critical exponent. The state of the art algorithm for
lattice simulations of QCD is the Hybrid Monte Carlo (HMC) \cite{Duane:1987de}.
With this algorithm the dynamical critical exponent
can be as large as $z\approx5$ \cite{Schaefer:2010hu}.
Employing open boundary conditions in time on can achieve
$z=2$ \cite{Luscher:2011kk}.

\section{Static quark potential}

The energy levels of a static quark and anti-quark pair at distance $r$
exhibits several of the fundamental properties of QCD and can be studied
by lattice QCD simulations. We denote the energy levels by
$V_n(r)$, $n=0,1,\ldots$. The ground state $V_{0}(r)\equiv V(r)$ is called
the static potential.

In pure $\SU{N}$ gauge theory the static potential is for small distances
$r<0.1\,{\rm fm}$ well described by a Coulomb potential $\propto 1/r$. 
At large distances $r\gg1\,{\rm fm}$ it becomes a confining potential which 
grows lineraly with the distance $r$. This linear growth extends to $r\to\infty$
in the pure gauge theory. In this case a flux tube forms between the static
quarks which is called (QCD) string.
When sea quarks are present the situation changes and the string breaks
at a distance $\rb\approx1.5\,\mathrm{fm}$ when the energy of the flux tube is
sufficient to create a sea quark and anti-quark pair which, together with
the static quarks, form a pair of static-light mesons.

\subsection{Continuum results}

Before we turn to the lattice, we mention some continuum results
which are known for the static potential.
At small distances $r\to0$ perturbation theory can be applied thanks to the
property of asymptotic freedom. At leading order the potential is determined
by the exchange of one gluon and is given by
\begin{equation}\label{e:onegluonexch}
V(r) \stackrel{r\to0}{\sim} -\CF\frac{g_0^2}{4\pi r}\,,\quad \CF=(N^2-1)/(2N) \,.
\end{equation}
The expansion of $V(r)$ in the $\msbar$ strong coupling 
$\as=\gbar^2_\msbar(1/r)/(4\pi)$ has been computed to two loops
\cite{Fischler:1977yf,Billoire:1979ih,Peter:1997me,Melles:2000dq}
and is now known to three loops
\cite{Brambilla:1999xf,Brambilla:1999qa,Smirnov:2009fh,Anzai:2009tm,Lee:2016cgz}
.
The potential contains an additive constant which originates from the
self-energy of the static quarks and diverges when the
ultra-violet cut-off is taken to infinity.
The static force
\begin{equation}
F(r) = \frac{{\rm d}V(r)}{{\rm d}r}
\end{equation}
is a renormalized quantity. A running strong coupling 
$\aqq(\mu) \equiv \gqq^2(\mu)/(4\pi)$
can be defined from the static force as
\begin{equation}
\aqq(\mu) = \frac{1}{\CF}r^2F(r) \,.
\end{equation}
The coupling $\gqq(\mu)$ runs with the energy scale $\mu=1/r$
according to the renormalization group equation
\begin{equation}\label{e:rgqq}
\mu \frac{{\rm d}}{{\rm d}\mu}\gqq(\mu) = \betaqq(\gqq(\mu)) \,,
\end{equation}
where the $\betaqq$ is the beta function.
The latter is known up to the 4-loop term
\begin{equation}\label{e:betaqq}
\betaqq(\gqq) =
-\gqq^3 [\sum_{n=0}^3 \bqq_n \gqq^{2n} +  \bqq_{3,l} \gqq^{6}\log(3 \gqq^2/(8\pi)) + \mathrm{O}({\gqq^8})] \,,
\end{equation}
where
$\bqq_0 = b_0=\frac{1}{(4\pi)^2}\left(11-\frac{2}{3}\nf\right)$ and
$\bqq_1 = b_1=\frac{1}{(4\pi)^4}\left(102-\frac{38}{3}\nf\right)$
are the universal coefficients. The non-universal coefficients 
$\bqq_2$, $\bqq_3$ and $\bqq_{3,l}$ are derived from the 3-loop
expression of the static potential in Appendix B of \cite{Donnellan:2010mx}.
The renormalization group equation \eq{e:rgqq} can be integrated in the form
\begin{eqnarray}
\Lambda_\mathrm{qq} &=&\mu\left(b_0\gqq^2\right)^{-b_1/(2b_0^2)} {\rm e}^{-1/(2b_0\gqq^2)} \nonumber \\
          && \times
           \exp \left\{-\int_0^{\gqq} {\rm d} x
          \left[\frac{1}{ \beta(x)}+\frac{1}{b_0x^3}-\frac{b_1}{b_0^2x}
          \right]
          \right\} \,, \label{e:Lambdaqq}
\end{eqnarray}
where $\Lambda_\mathrm{qq}$ is the $\Lambda$ parameter in the $\mathrm{qq}$ 
scheme. The latter can be computed from the $\Lambda$ parameter in the
$\msbar$ scheme. Given a value for $\mu/\Lambda_\mathrm{qq}$,
the numerical $n$-loop solution $\gqq=\gqq(\mu/\Lambda_\mathrm{qq})$
to \eq{e:Lambdaqq} is obtained by truncating 
the $\betaqq$ function in the integrand after the $\bqq_{n-1}$ term and 
solving for $\gqq$.

As $r\to\infty$ the potential in the pure gauge theory can be computed using 
an effective bosonic string theory 
\cite{Nambu:1978bd,Luscher:1980fr,Luscher:1980ac,Luscher:2004ib}.
The string describes a flux tube in $d-1$ dimensions joining the static sources 
and fluctuating in $d-2$ transverse directions
\begin{equation}\label{e:string}
V(r) = \sigma r + \mu + \frac{\gamma}{r} + O(1/r^3) \,,
\end{equation}
where $\sigma$ is the string tension, $\mu$ a mass parameter and the
coefficient $\gamma=-\pi(d-2)/24$ depends only on $d$.
The width of the string increases logarithmically in the distance $r$ 
\cite{Luscher:1980iy}.

\subsection{The static potential on the lattice}

\begin{figure}[t]\centering
\includegraphics[width=.25\textwidth]{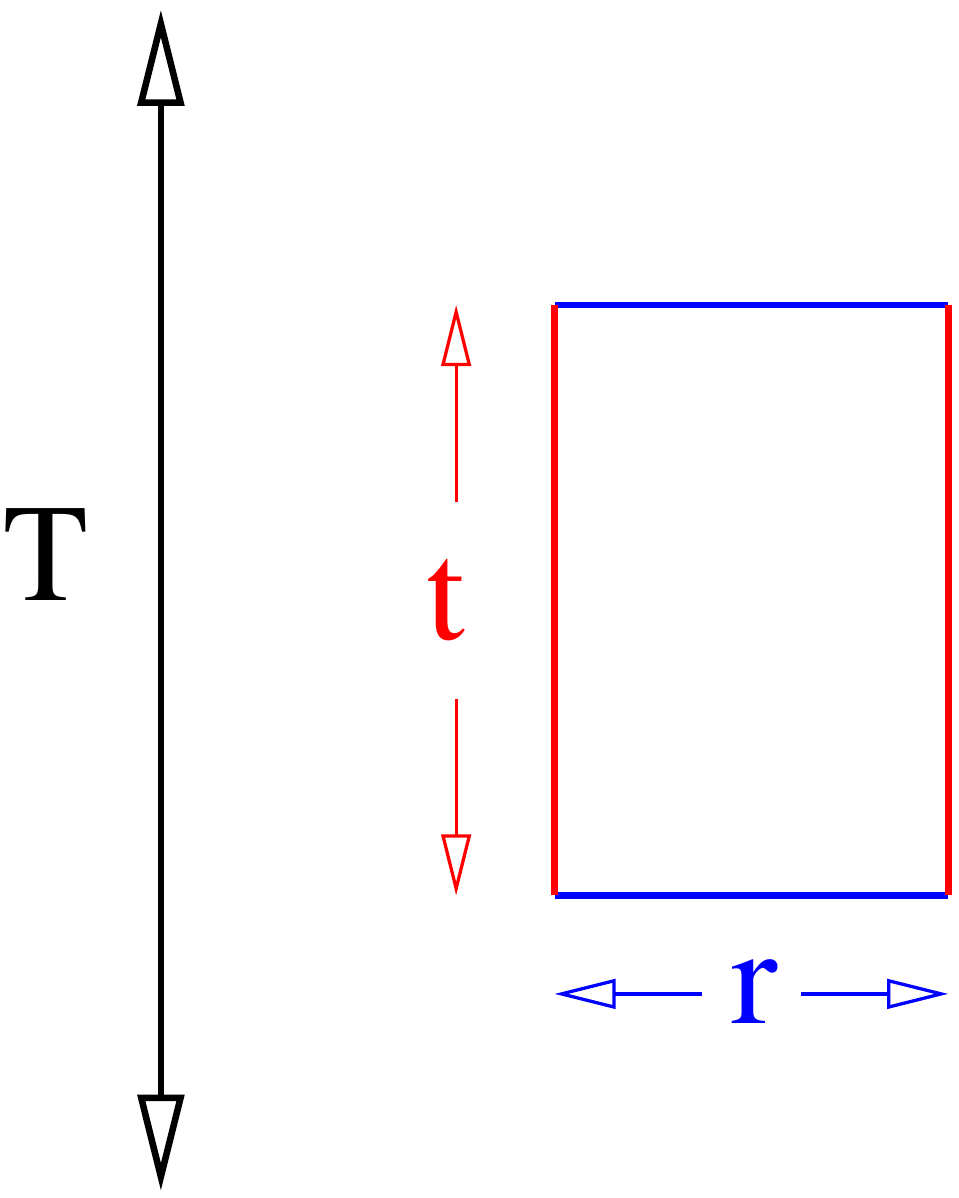}
\hspace{2cm}
\includegraphics[width=.27\textwidth]{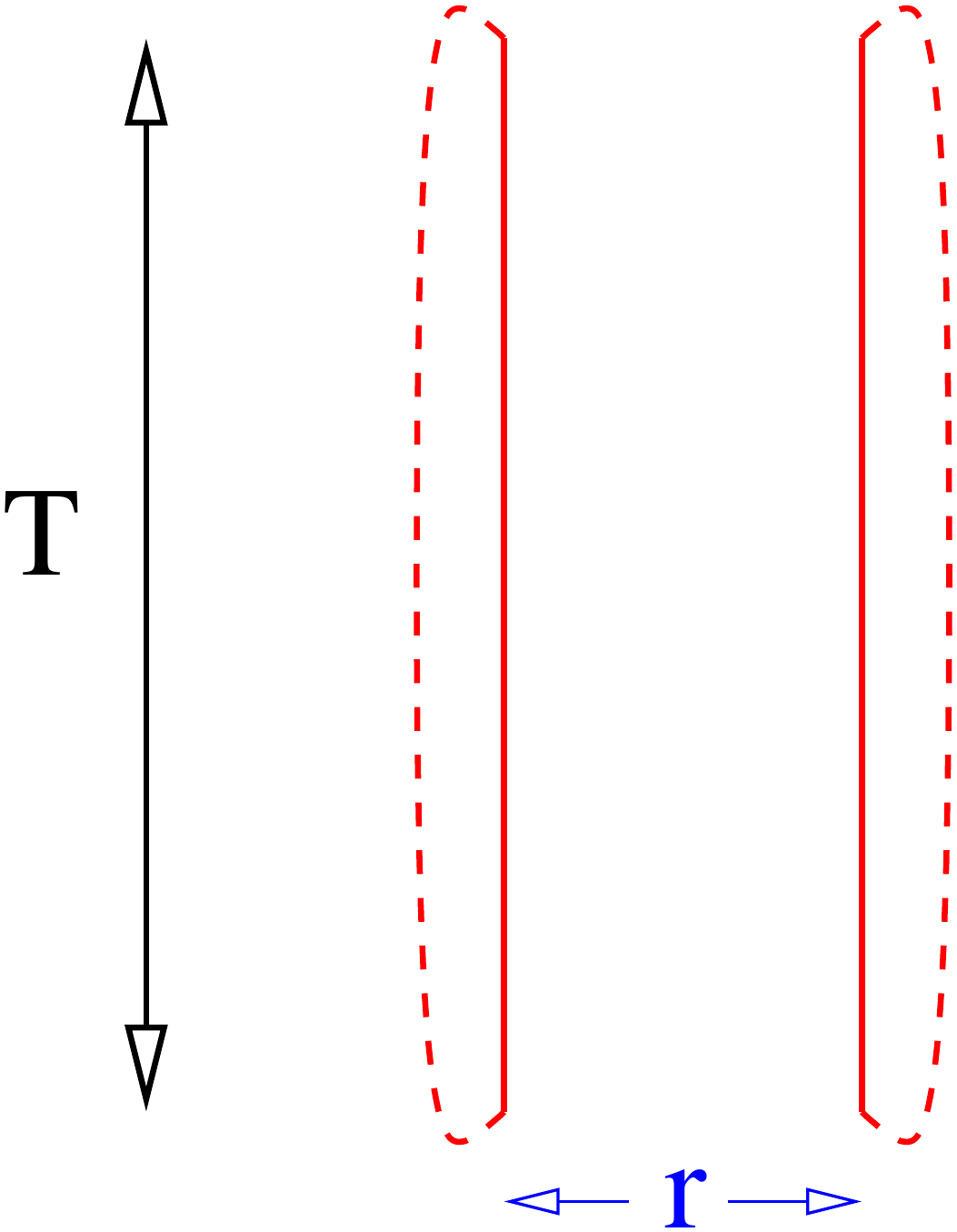}
\caption{Wilson loop (left) and correlator of Polyakov loops (right).}
  \label{f:loops}
\end{figure}

On the lattice the static potential can be extracted from Wilson loops, 
which are shown on the left of \fig{f:loops}.
A planar Wilson loop is defined as the trace of the product of gauge links 
along a rectangle of size $r\times t$
\begin{equation}\label{e:Wloop}
W(r,t) = 
\tr\left\{ P(0,\underline{0};0,r\hat{k})
P(0,r\hat{k};t,r\hat{k})
P^\dagger(t,\underline{0};t,r\hat{k})
P^\dagger(0,\underline{0};t,\underline{0}) \right\} \,.
\end{equation}
Here $P(x_0,\underline{0};x_0,r\hat{k})$ is the product of spatial links 
joining the static sources at time $x_0$. It represents a string-like state.
$P^\dagger(0,\underline{x};t,\underline{x})$ is the product of temporal links at position 
$\underline{x}$. It represents the propagator of the static quark.
Assuming that the theory has an (effective) transfer matrix
the expectation value of the Wilson loop has the spectral representation
\begin{eqnarray}\label{e:Wloopspec}
\ev{W(r,t)} & \stackrel{t\to\infty}{\sim} &
\sum_n c_nc_n^* {\rm e}^{-V_n(r)t} \,.
\end{eqnarray}
This representation is for example derived explicitely in Appendix B of
\cite{Knechtli:1999tw}.
In pure gauge theory for large values of $r$ and $t$ one expects an area law
$\ev{W(r,t)}\sim{\rm e}^{-\sigma r t}$.

The standard way to extract the energy levels $V_n$ from \eq{e:Wloopspec}
is to use some technique of
gauge link smearing to replace the spatial lines in the Wilson loops by
$P_i(x_0,\vec{0};x_0,r\hat{k})$, where $i$ is the smearing index.
One then has a correlation matrix $\ev{W_{ij}(r,t)}$ (the smearing index $i$
refers to time $0$ and $j$ to time $t$. 
For fixed distance $r$ the correlation matrix 
$C_{ij}(t)=\ev{W_{ij}(r,t)}$ is used to build the generalized eigenvalue problem
\begin{equation}\label{e:gevp}
C(t) v_n(t,t_0) = \lambda_n(t,t_0) C(t_0) v_n(t,t_0) \,,\quad
\lambda_n> \lambda_{n+1}
\end{equation}
The energy levels $V_n$ can be computed as \cite{Luscher:1990ck}
\begin{equation}\label{e:potentials}
V_n(r) = -\frac{1}{a}\ln\left\{
\frac{\lambda_n(t+a,t_0)}{\lambda_n(t,t_0)}\right\} + \epsilon_n(t,t_0)
\end{equation}
with exponential corrections $\epsilon_n={\rm O}({\rm e}^{-\Delta E_nt})$.
We refer to \cite{Blossier:2009kd} for a derivation of the corrections 
$\Delta E_n$.

The static potential can also be extracted from the spatial
correlator of temporal Polyakov loops, which is
shown on the right of \fig{f:loops}:
\begin{equation}
P(x)^*P(y)|_{y=x+r\hat{k}} \,,
\end{equation}
where the Polyakov loop is defined as the trace of the path through $x$ 
which winds once around the time axis
$P(x)=\tr\{\gauge{\mu}{x}\gauge{\mu}{x+a\hat{\mu}}\ldots\gauge{\mu}{x+(T-a)\hat{\mu}}\}|_{\mu=0}$.
The static potential is obtained as \cite{Luscher:2002qv}
\begin{eqnarray}
V(r) & \stackrel{T\to\infty}{\sim} & -\frac{1}{T}\ln\ev{P(x)^*P(y}|_{y=x+r\hat{k}}
\,.
\end{eqnarray}
Using the Polyakov loop correlator requires simulations of lattices with
different temporal extents $T$ to take the limit $T\to\infty$.

Wilson loops are not good operators when it comes to compute the
static potential in QCD with dynamical quarks (in the fundamental
representation of $\SU{3}$) at large distances $r$. As we mentioned above 
at large $r$
due to string breaking the ground state potential is the energy of
two static-light mesons which results in a flattening of the static potential.
As it was shown in previous studies in the case of the $\SU{2}$ Higgs
model \cite{Knechtli:1998gf,Knechtli:2000df,Philipsen:1998de}
and then in QCD \cite{Bali:2005fu},
in order to see string breaking and the flattening of the potential one has
to construct a matrix correlation which includes Wilson loops, off-diagonal
matrix elements describing the transition between a
string-like state and a state made of two static-light mesons and
a diagonal element describing only two static-light mesons.
Schematically the matrix is represented as
\begin{equation}\label{e:matrixsb}
\left(\begin{array}{rl}
\quad \www&\quad \sqrt{\nf}\wwb\\&\\
\sqrt{\nf}\wbw&\quad -\nf\wbbc+\wbbd\end{array}\right) \,,
\end{equation}
where $\nf$ degenerate dynamical quarks are assumed and 
the diagrams are from \cite{Bali:2005fu}.

Due to confinement, the Wilson loop has a signal which decays exponentially
with the area of the loop:
$\ev{W(r,t)} \approx \exp(-\sigma r t)$. This is the result of cancellations
in the average between positive and negative values of the loop.
The variance of the Wilson loop is instead dominated by
$\ev{W(r,t)^2}$ which is approximately a constant (being the average of
positive values). Therefore
the signal-to-noise ratio of Wilson loops decays exponentially with the area
of the loop. The same problem arises with Polyakov loops.
The deterioration of the quality of the signal can be
tamed by the following techniques:
\begin{itemize}
\item
In pure $\SU{N}$ theory, an exponential reduction 
with the temporal extent $t$ of the error of Wilson loops 
can be achieved by using the
one-link integral to replace the temporal link \cite{Parisi:1983hm}.
An even more effective technique to reduce exponentially the error
of the Polyakov loop correlator
is the multilevel algorithm \cite{Luscher:2001up}. 
These methods are not applicable with fermions.
\item
With fermions, HYP smearing \cite{Hasenfratz:2001hp} can be used to smear
the temporal gauge links in the Wilson loops \cite{Donnellan:2010mx}.
This minimizes the self-energy contribution of the static quarks and
therefore improves the signal-to-noise ratio \cite{DellaMorte:2005nwx}.
Still the signal-to-noise ratio falls off exponentially.
\end{itemize}
Smearing is also applied to the quark fields in the correlation matrix
\eq{e:matrixsb} to improve the overlap with the physical states.
An efficient quark smearing technique is distillation \cite{Peardon:2009gh}.

\subsection{Lattice results}
\begin{figure}[t]\centering
\includegraphics[width=.45\textwidth]{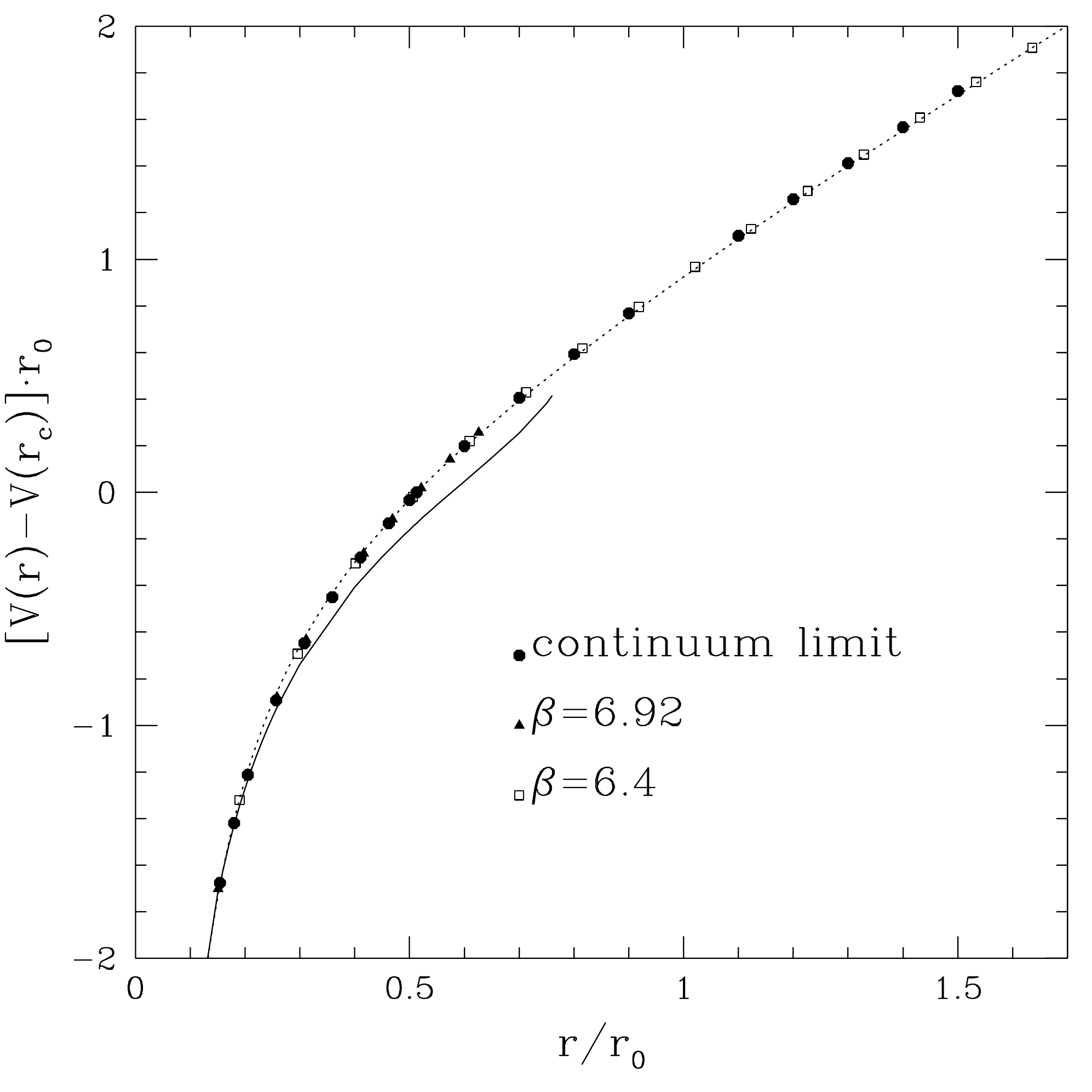}
\hspace{0.1cm}
\includegraphics[width=.5\textwidth]{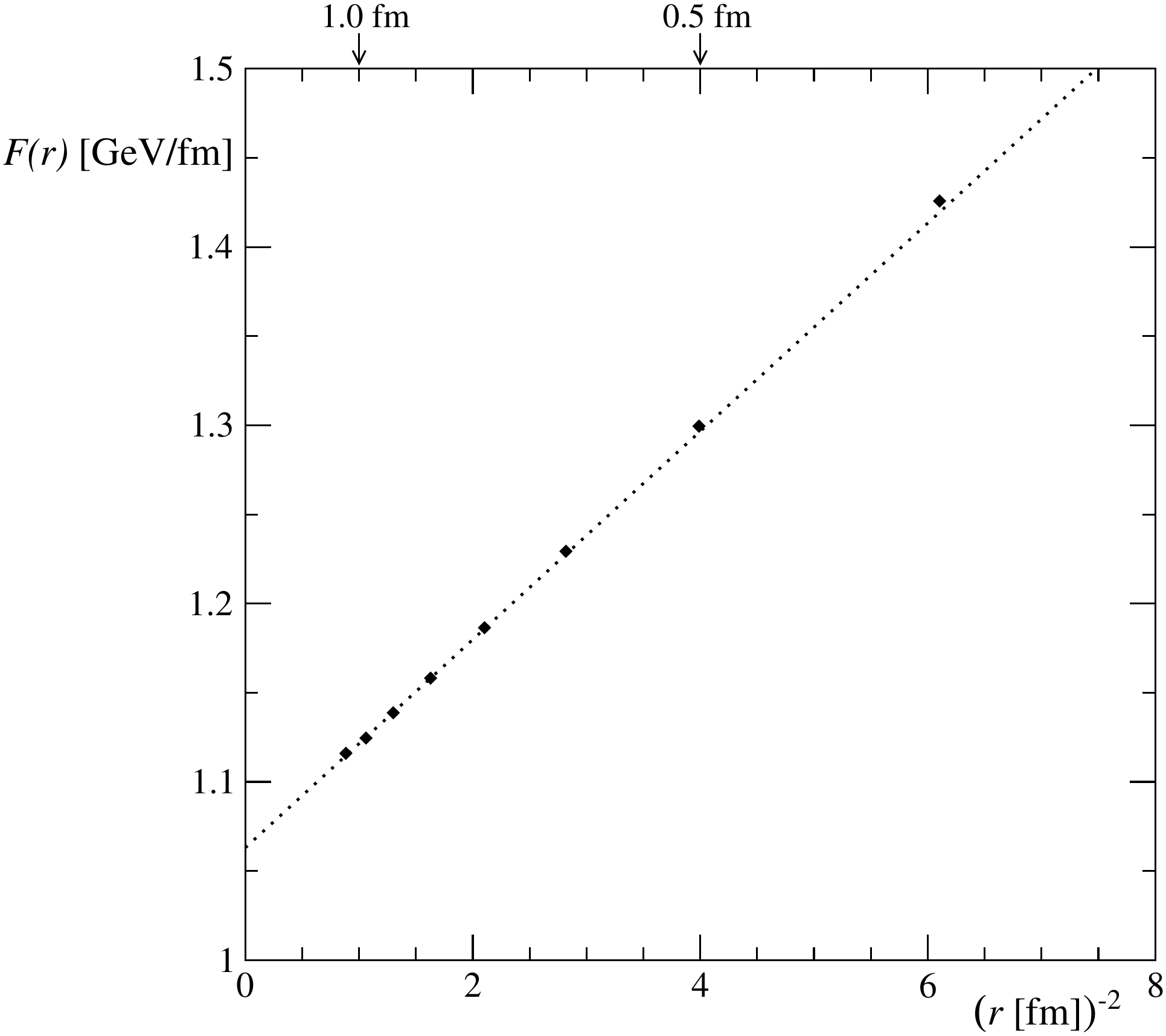}
\caption{Left: the static potential in the pure gauge theory, from \cite{Necco:2001xg}. Right: the static force in the pure gauge theory, from \cite{Luscher:2002qv}.}
  \label{f:potquenched}
\end{figure}

The static potential can be used to define a physical scale by solving
the following equation for the static force
\begin{equation}\label{e:rc}
r^2 F(r)|_{r=r(c)} = c \,.
\end{equation}
Specifying a value for the constant $c$ and solving for $r=r(c)$ leads to a
value $r(c)/a$ in lattice units. If $r(c)$ is known in physical units of
$\mathrm{fm}$ (fermi) the lattice spacing can be determined. Taking
$c=1.65$ leads to the Sommer scale $r_0=r(1.65)$ which has a value
of about $0.5\,\mathrm{fm}$ in QCD \cite{Sommer:1993ce}. The physical value
can be determined by comparing to phenomenological potential models.
Alternatively it can be extracted by computing the product with another
scale, for example with the kaon decay constant $r_0f_{\rm K}$ and using
the physical value of $f_{\rm K}$, see \cite{Fritzsch:2012wq}.
Other choices for $c$ in \eq{e:rc} are $r_1=r(1.0)$ \cite{Bernard:2000gd}
and $r_c=r(0.65)$ \cite{Necco:2001xg}.
\begin{figure}[t]\centering
\includegraphics[width=.45\textwidth]{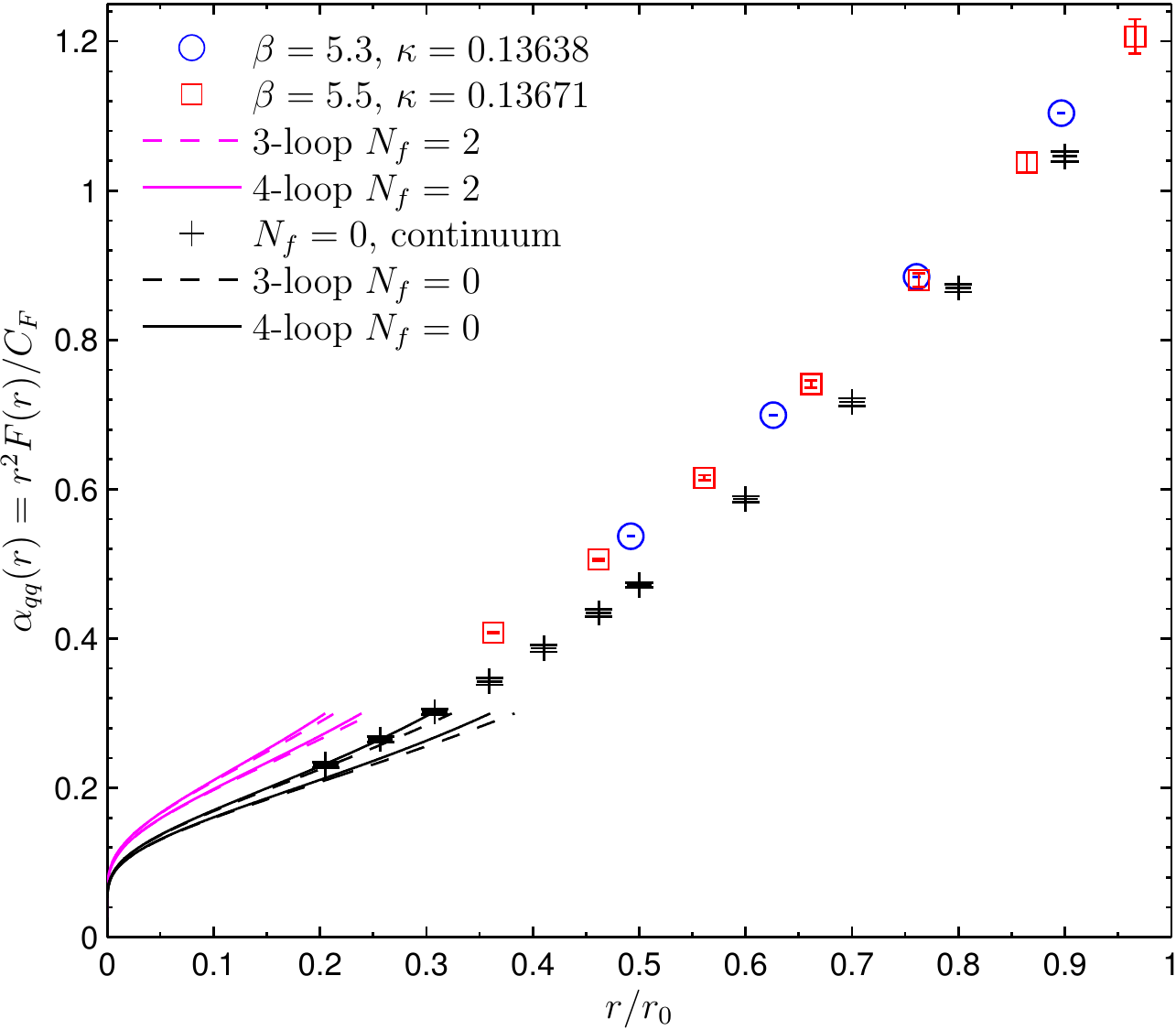}
\hspace{0.5cm}
\includegraphics[width=.46\textwidth]{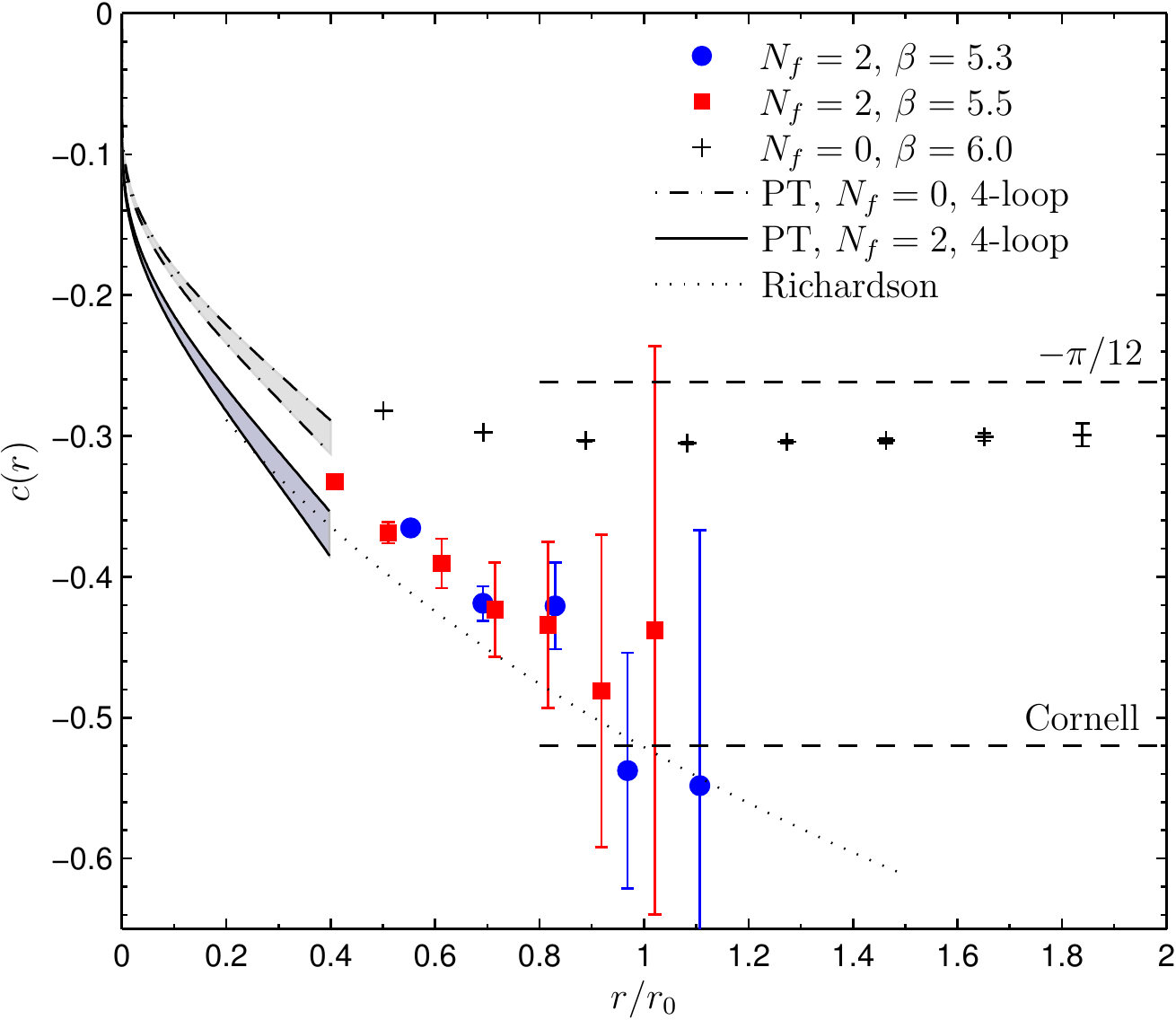}
\caption{Left: the running coupling $\aqqr$ in the $\nf=2$ theory compared to
perturbation theory and $\nf=0$ data from \cite{Necco:2001xg}. Right: the slope of the static force $c(r)$, which is proportional to a running coupling, compared to pertutbation theory, potential models and $\nf=0$ data from \cite{Luscher:2002qv}. The plots are taken from \cite{Leder:2011pz}.}
  \label{f:couplings}
\end{figure}

In this section we present a selection of results about the static potential
obtained from lattice simulations which illustrate fundamental properties
of QCD.
In \fig{f:potquenched} we show results for the static potential (left)
and the static force (right) in the pure gauge theory.
The left plot is taken from \cite{Necco:2001xg} and shows the static potential
after taking the continuum limit (black circles). At short
distances it is compared to the perturbative expansion (continued line).
The latter is computed from the static force
by integrating the renormalization group 
equation for $\aqq$ as in \eq{e:Lambdaqq} with the 3-loop $\beta$-function
(i.e. using the coefficients of the $\betaqq$ function up to and including
$\bqq_{2}$ in \eq{e:betaqq}).
The dotted line is the parameter free prediction from the bosonic string
in \eq{e:string}, where the string tension is fixed by 
$\sigma r_0^2=1.65-\pi/12$.
One sees that the bosonic string model agrees very well with the data 
for the static potential already at distances around $r_0$.
This is confirmed by the right plot in \fig{f:potquenched}, which is
taken from \cite{Luscher:2002qv} and shows the static force as a
function of $1/r^2$. 
The bosonic string predicts $F(r)=\sigma-\gamma/r^2+{\rm O}(1/r^4)$.
Indeed the data are
well approximated by a linear function in
$1/r^2$ (dotted line, fitted to the four points at the largest distances)
as soon as $r\ge r_0\simeq0.5\,\mathrm{fm}$. Still there is a curvature in
the data which can be measured by computing the slope
\begin{equation}\label{e:slope}
c(r)=\frac{1}{2}r^3F^\prime(r)\,,
\end{equation}
which is shown in the right plot of \fig{f:couplings}. The slope $c(r)$ can be used to define a running
coupling $\ac=\gc^2/(4\pi)$ for small $r$
\begin{equation}\label{e:ac}
\ac(\mu) = -\frac{1}{\CF}c(r)
\end{equation}
at the renormalization $\mu=1/r$. The beta function for the coupling
$\ac$ can be found in Appendix B of \cite{Donnellan:2010mx}.

The coupling $\aqq$ and the slope $c(r)$ have been measured in the theory
with $\nf=2$ dynamical fermions. The results are shown in \fig{f:couplings}.
We used the ensembles of gauge configurations
labelled ``F7'' and ``O7'' in \cite{Fritzsch:2012wq} which were
generated by CLS (Coordinated Lattice Simulations consortium) using the Wilson 
gauge action and $\nf=2$ flavors of O($a$) improved Wilson quarks. The
lattice spacings are $a=0.066\,\mathrm{fm}$ (blue circles) and 
$a=0.049\,\mathrm{fm}$ (red squares) and the quark mass corresponds to
a pion mass $m_\pi=270\,\mathrm{MeV}$.
In \fig{f:couplings} we also plot the perturbative curves obtained 
from \eq{e:Lambdaqq} (and a similarly for the coupling $\gc$).
The $\Lambda$ parameters are known from \cite{Capitani:1998mq} for the
$\nf=0$ theory and from \cite{Fritzsch:2012wq} for the $\nf=2$ theory.
The band of the perturbative curves reflects the
uncertainty of the $\Lambda$ parameter.
A quantitative comparison to the perturbative curves requires
a careful continuum limit and, certainly for the $\nf=2$ theory,
smaller lattice spacings to reach small enough distances $r$.
In the right plot about $c(r)$ in \fig{f:couplings} we compare the
$\nf=2$ data to the value $c=-0.52$ 
 (dashed line)\footnote{
This is approximately twice the asymptotic value $c(r=\infty)=-\pi/12$
in the pure gauge theory.} that it takes in the
phenomenological Cornell potential \cite{Eichten:1979ms} and to
the curve (dotted) derived from the Richardson potential
\cite{Richardson:1978bt}.
The slope $c$ is an interesting but difficult quantity for holographic QCD models \cite{Giataganas:2011nz}.
\begin{figure}[t]\centering
\includegraphics[width=.42\textwidth]{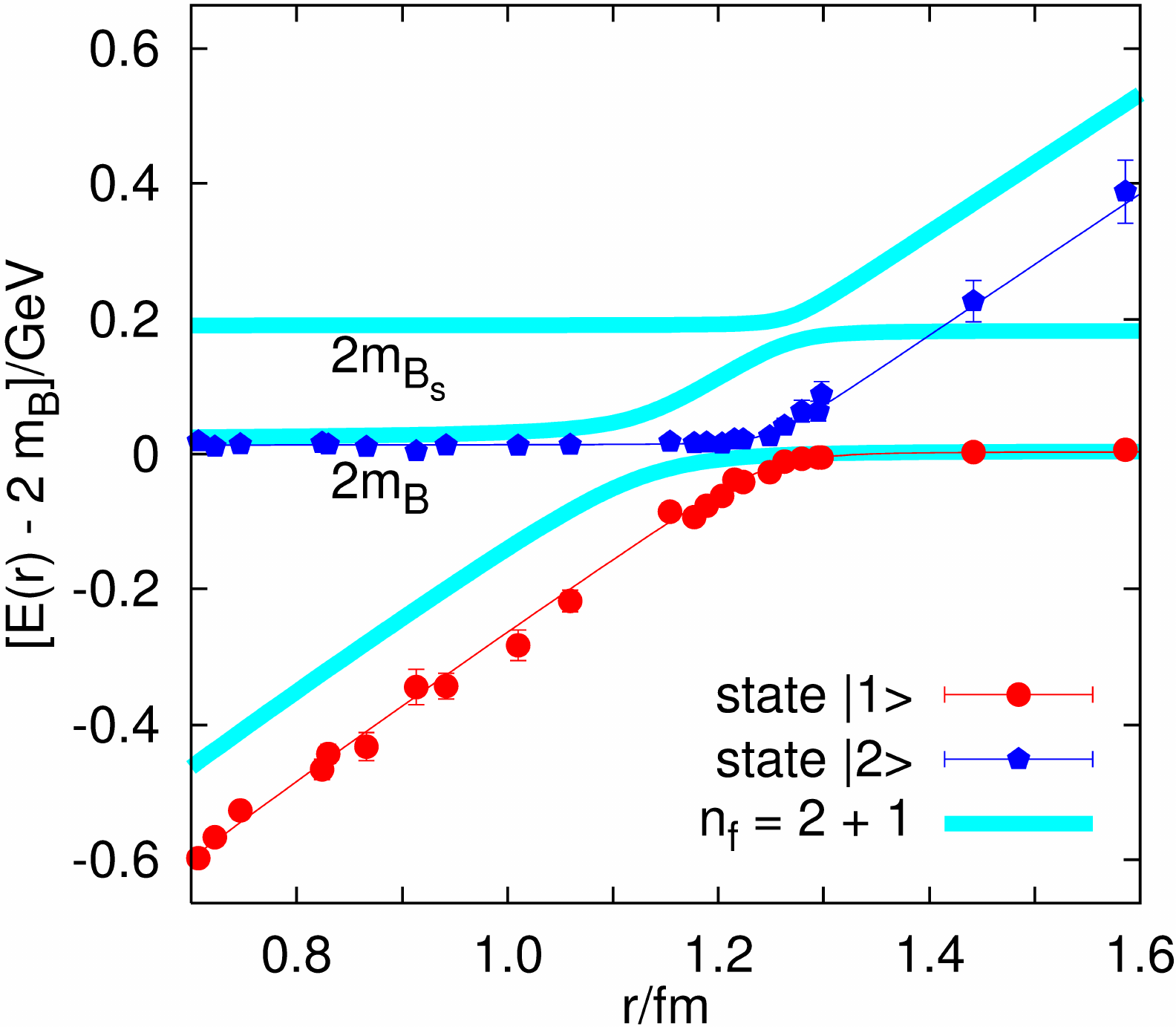}
\hspace{0.1cm}
\includegraphics[width=.53\textwidth]{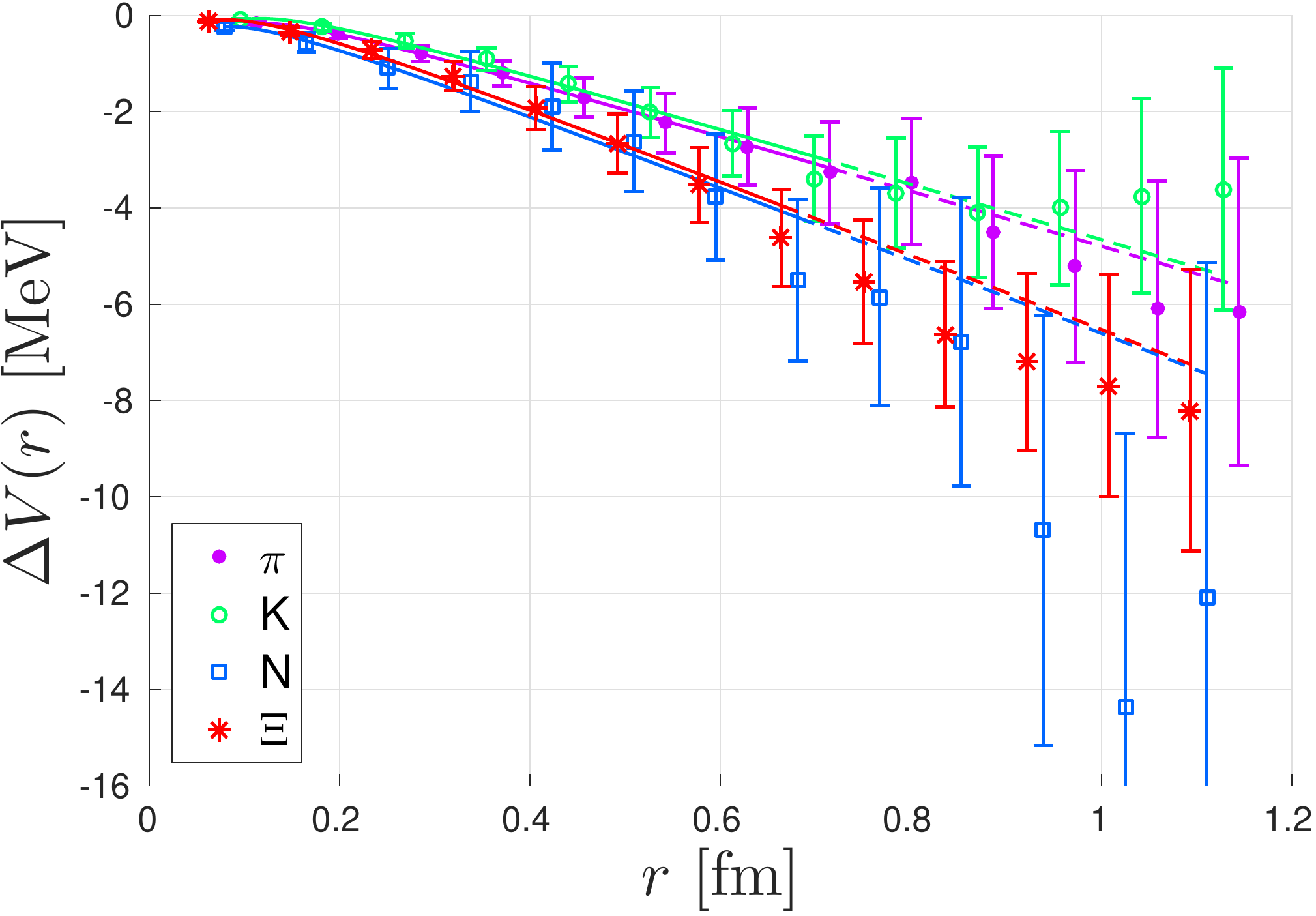}
\caption{Left plot: the static potential (red circles) and its first excited state (blue pentagrams), from \cite{Bali:2005fu}. Right plot: the modification of the static potential in the presence of a hadrons, from \cite{Alberti:2016dru}.}
  \label{f:sb_hadroquark}
\end{figure}

The energy levels for the ground state and the first excited state potentials
in QCD with $\nf=2$ dynamical quarks have been calculated in \cite{Bali:2005fu}
from the correlation matrix \eq{e:matrixsb}.
The simulations are done at a lattice spacing $a=0.083\,\mathrm{fm}$ and pion
mass $m_\pi=400\,\mathrm{MeV}$ and the results are shown in the left plot
of \fig{f:sb_hadroquark}. The string breaking distance was calculated to be
$\rb/r_0\approx2.5$. The cyan bands are a qualitative sketch of the expected 
behavior with $\nf=2+1$ dynamical quarks, where a second breaking happens. 
Calculations of string breaking with $\nf=2+1$ are under way \cite{Koch:2015qxr}. The continuum limit and the dependence on the quark mass of string breaking
are still open issues.
String breaking is a fundamental aspect of QCD with dynamical quarks. It 
  provides input for phenomenological potential models which can shed light 
  on the nature of recently-discovered exotic heavy-flavor hadrons, see
\cite{Oncala:2017hop} for a recent discussion.

The static potential is modified by the presence of a hadron.
The difference $\Delta V_H=V_H-V_0$ of the static potential in the presence
of a hadron $V_H$ and the static potential in the vacuum $V_0$ 
can be computed from the correlator 
of a Wilson loop $W(r,t)$ and a hadron two-point function $C_{H,\mathrm{2pt}}$
defined as \cite{Alberti:2016dru}
\begin{equation}
C_H(r, \delta t, t) =
\frac{\langle W(r,t)C_{H,\mathrm{2pt}}(t+2\delta t)\rangle}{\langle W(r,t)\rangle
\langle C_{H,\mathrm{2pt}}(t+2\delta t)\rangle} \,.
\end{equation}
The argument $\delta t$ is the distance in time between the source of the hadron and the Wilson loop and it is also the same distance between the latter and the sink of the hadron (i.e. the hadron
propagates over a time $t+2\delta t$).
The potential difference is computed as $\Delta V_H = -\lim_{t\rightarrow\infty}\frac{{\rm d}}{{\rm d}t}\ln[C_H(r, \delta t, t)]$ and extrapolating $\delta t\to\infty$. 
The right plot of \fig{f:sb_hadroquark} presents the results for the
shift $\Delta V_H$ for the pion $\pi$, the kaon $K$, the nucleon $N$ and the 
cascade $\Xi$. The calculation was done using the $\nf=2+1$ ensemble of gauge configurations labelled ``C101'' generated by CLS \cite{Bruno:2014jqa}.
The lattice spacing is $a=0.0854(15)\,\mathrm{fm}$ and the kaon and pion
masses are $m_\pi\approx223\,\mathrm{MeV}$ and
$m_\mathrm{K}\approx476\mathrm{MeV}$ respectively.
The potential shift is negative and of similar size for all hadrons 
investigated. The effect is small, it is 
$-\Delta V_H\approx2$--$3\,\mathrm{MeV}$
at distance $r=0.5\,\mathrm{fm}\simeq r_0$. 
Solving the Schr{\"o}dinger equation with the modified potential
yields a stronger binding of charmonium states ($c\bar{c}$). 
Their masses decrease by few MeV like the binding of deuterium.
The modification of the static potential in the presence of a hadron is
a test of the idea of hadro-quarkonia \cite{Dubynskiy:2008mq}
which provides a possible interpretation
for the candidates of a penta-quark state ($c\bar{c}uud$) reported in
\cite{Aaij:2015tga,Aaij:2016phn}.

\section{Conclusions}

More than 40 years after its invention, lattice QCD has
developed into an active field of research connecting physics, mathematics and 
informatics. As it was shown in this lecture using the example of the static 
potential, there are still important open questions to be answered.

{\bf Acknowledgments.} I thank Peter Weisz for his help
in providing me with his lecture at the Corfu Summer School 2014
and Tomasz Korzec for his valuable comments on the manuscript.
I acknowledge financial support from the Program for European Joint
Doctorates HPC-LEAP (High Performance Computing in Life sciences,
Engineering and Physics).

\end{document}